\documentclass{ws-p8-50x6-00}

\usepackage{epsfig}

\def\Id{{\rm 1\kern-.3em I}}
\def\EPJC{{\em Eur. Phys. J.} C}

\def\NPA{{\em Nucl. Phys.} A}
\def\rightnote{{\it Proceedings of NSTAR 2001, Mainz; submitted to {\bf World Scientific}}}%

\begin{document}

\title{
Parity Doublets from a\\
Relativistic Quark Model
}

\author{
Ulrich L\"oring, Bernard Metsch
}

\address{
Institut f\"ur Theoretische Kernphysik der Universit\"at Bonn,\\
Nu{\ss}allee 14-16, D-53115 Bonn, GERMANY\\
E-mail: loering@itkp.uni-bonn.de,
metsch@itkp.uni-bonn.de
}

\maketitle

\abstracts{
The $N$-- and the $\Lambda$--excitation spectrum
exhibit parity doublets, {\it i.e.} states of the same
spin but with opposite parity being almost degenerate in mass. It is
shown that in a relativistic quark model with instantaneous
interaction kernels, where confinement is
implemented by a linearly rising potential and the major mass
splittings are generated from an interaction based on instanton
effects, this degeneracy occurs quite naturally: Once the parameters
of the instanton induced interaction are fixed to reproduce the ground
state octet-decuplet splittings, some states are selectively lowered
to a position which is degenerate with states of opposite
parity. Some observable consequences are briefly discussed.
}

A glance at the nucleon-- and $\Lambda$-- excitation spectrum\cite{PDG00}
reveals a conspicuous degeneracy of some states with the same spin and
opposite parity. Prominent examples are
$N\frac{5}{2}^+\!(1680)$--$N\frac{5}{2}^-\!(1675)$, 
$N\frac{9}{2}^+\!(2220)$--$N\frac{9}{2}^-\!(2250)$,
$\Lambda\frac{5}{2}^+\!(1820)$--$\Lambda\frac{5}{2}^-\!(1830)$\,. Although
one might also regard 
$\Delta\frac{5}{2}^+\!(1905)$--$\Delta\frac{5}{2}^-\!(1930)$ and 
$\Delta\frac{9}{2}^+\!(2300)$--$\Delta\frac{9}{2}^-\!(2400)$ as parity
partners, the situation seems less clear: for the first because of the
nearby $\Delta\frac{5}{2}^+\!(2000)$-resonance and for the second because of
the relatively large splitting. In the $\Sigma$--spectrum no
clear indications of parity doublets is found. In the literature these
observations have been related to a phase transition from the
Nambu--Goldstone mode of chiral
symmetry to the Wigner--Weyl mode in the upper part of the baryon
spectrum\cite{Glozman:2000tk,Cohen:2001gb,Kirchbach:2001hj}\,. 
In the present contribution we will show how this feature can be
understood in the context of a (relativistic) constituent quark model
on the basis of the quark {\em dynamics}, where the major spin-dependent 
mass splittings are induced by instanton effects ('t Hooft's force).

Our relativistically covariant constituent quark model is based on the
Bethe-Salpeter equation for three-quark bound states 
with instantaneous interaction
kernels. The details of the model are described elsewhere\cite{Loring:2001kvA,Loring:2001kvB,Loring:2001kvC}.
Quarks are assumed to possess an effective constituent mass and
confinement is implemented by a 
linearly rising 3-body string potential with a Dirac structure,
which is a combination of scalar and time-like vector structures 
chosen such that unwanted spin-orbit effects are minimized.
The major spin-dependent mass splittings are generated by a flavor dependent
2-particle interaction, which is motivated by instanton effects.
This force affects flavor-antisymmetric $qq$-pairs only, and
consequently this interaction does not act on flavor symmetric states,
such as the $\Delta$-resonances, which are thus determined by the
dynamics of the confinement potential alone. Accordingly, the
constituent quark masses and the confinement parameters were
determined by a fit to the spectrum in this
sector\cite{Loring:2001kvB}.

\begin{figure}[!htb]
\begin{center}
	\epsfxsize=0.7\textwidth
	\epsfbox{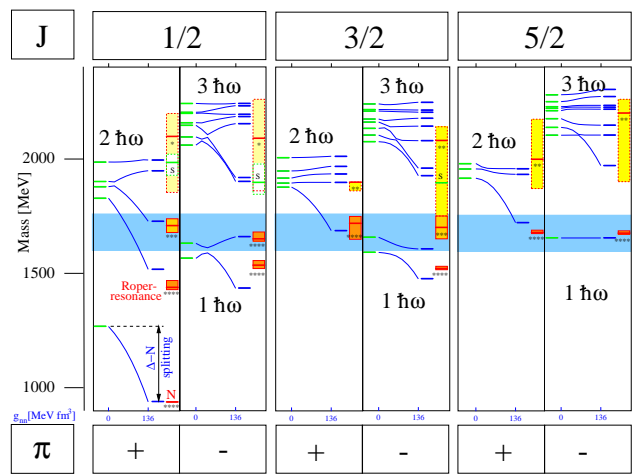}
\end{center}
\vspace*{-4mm}
	\caption{
	Parity doublets for low lying nucleon resonances. The left
 	side of each column shows the spectrum calculated with the
 	confinement interaction alone. The curve shows the effect of
 	the instanton induced interaction with increasing coupling $g_{nn}$
 	of the nonstrange-nonstrange quark interaction	
 	until the $N$--$\Delta$ splitting is reproduced. This result
	is compared to the experimental spectrum{\protect \cite{PDG00}}, right part of each
 	column, where uncertainties in the
 	resonance position are indicated by boxes.
\label{fig:nuclow}
}
\vspace*{-3mm}
\end{figure}

\begin{figure}[!htb]
	\epsfxsize=\textwidth
	\epsfbox{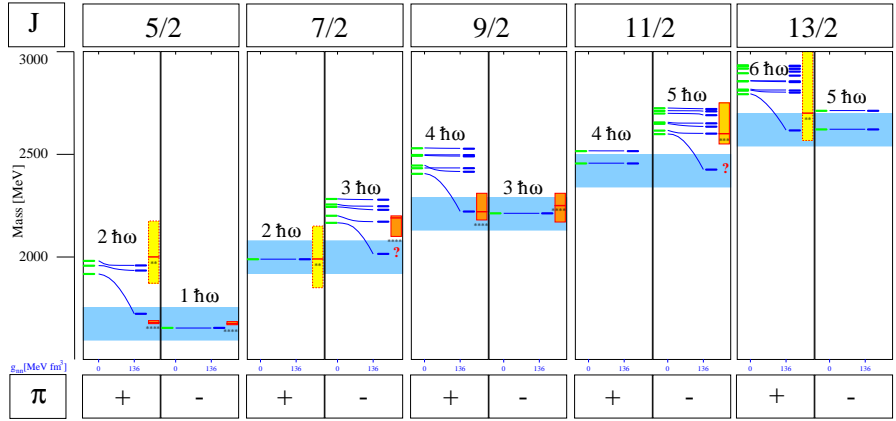}
\vspace*{-7mm}
	\caption{
	Parity doublets for higher lying nucleon resonances. 
	See also caption to Fig.~\ref{fig:nuclow}\,.
\label{fig:nuchigh}
}
\vspace*{-8mm}
\end{figure}

The residual instanton induced interaction does act
on particular flavor octet states. Once the strengths of this interaction have
been adjusted to account for the ground state nucleon and
$\Lambda$-mass we find that in fact one can describe the major
spin-dependent mass splittings in the nucleon spectrum quite well\cite{Loring:2001kvB}, see
Fig.~\ref{fig:nuclow}\,.
In particular one finds that in this manner the Roper resonance can be
accounted for quite naturally. Moreover one finds a selective lowering
of those substates of a major oscillator shell (which in spite of the
linear confinement adopted here still provides an adequate
classification of states with confinement alone) which contain so
called scalar diquarks, {\it i.e.} quark pairs with trivial spin and
angular momentum. This is found in particular for the highest spin
states in a given oscillator shell $N$, see Fig.~\ref{fig:nuchigh}\,.
For given $N\hbar\omega$ the maximum total angular momentum for a
state containing such a scalar diquark is $J = (L_{\mbox{\footnotesize
max}}=N) + \frac{1}{2}$\,. 't Hooft's force lowers this state enough
to become almost degenerate with the unaffected spin-quartet state of
the oscillator shell with $N-1$, which has opposite parity but the
same total angular momentum: $J = (L_{\mbox{\footnotesize max}}=N-1) +
\frac{3}{2}$. In this way patterns of approximate parity doublets for
{\it all} lowest excitations in the sectors $J=\frac{5}{2}$ to
$J=\frac{13}{2}$ are formed {\it systematically}.  In the
$N\frac{5}{2}^\pm$ and $N\frac{9}{2}^\pm$ sectors this scenario is
nicely confirmed experimentally by the well-established parity
doublets $N\frac{5}{2}^+(1680)$--$N\frac{5}{2}^-(1675)$ and
$N\frac{9}{2}^+(2220)$--$N\frac{9}{2}^-(2250)$. In the $N\frac{7}{2}$
sector, however, the present experimental findings seem to deviate
from such a parity doubling structure due to the rather highly
determined resonance position of the $N\frac{7}{2}^-(2190)$.  Although
this state is given a four-star rating\cite{PDG00} an investigation of
this sector with new experimental facilities such as the CLAS
detector at CEBAF (JLab) or the Crystal Barrel detector at ELSA (Bonn)
would be highly desirable. The same mechanism explains approximate
parity doublet structures also for states with lower angular momentum
as {\it e.g.} the $N^*$ doublets in the second resonance region around
$\sim1700$ MeV with spins $J^\pi=\frac{1}{2}^\pm$, $\frac{3}{2}^\pm$, and
$\frac{5}{2}^\pm$ (see fig.~\ref{fig:nuclow}).

Observable consequences of this parity doubling scenario should
manifest in a different shape of electromagnetic $p\gamma^*\rightarrow
N^*$ transition form factors of both members of a doublet due to their
significantly different internal strutures.  Fig. \ref{fig:formfact}
shows as an example the magnetic multipole $\gamma^*p\rightarrow
N\frac{5}{2}^+(1680)$ and $\gamma^*p\rightarrow N\frac{5}{2}^-(1675)$
transition form factors: 
\begin{figure}[t]
\begin{center}
\epsfxsize=0.6\textwidth
\epsfbox{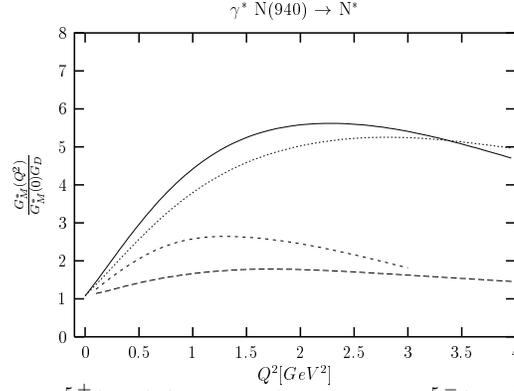}
\end{center}
\vspace*{-7mm}
\caption{The $\gamma^*p\rightarrow N\frac{5}{2}^+(1680)$ (dotted line) and
$\gamma^*p\rightarrow N\frac{5}{2}^-(1675)$ (short dashed line) transition form factors
$G^*_M(Q^2)$ divided by the dipole form $G_D(Q^2)$ and normalized to
their threshold values $G^*_M(0)$. For comparison also the $\gamma^*p\rightarrow N\frac{1}{2}^-(1535)$
(solid line) and $\gamma^*p\rightarrow \Delta\frac{3}{2}^+(1232)$ (dashed line) are shown.}
\label{fig:formfact}
\vspace*{-7mm}
\end{figure} 
That member of the doublet, which is affected by 't Hooft's
force ($N\frac{5}{2}^+$), exhibits a rather strong scalar diquark correlation
and thus its structure should be more compact compared to its unaffected
doublet partner ($N\frac{5}{2}^-$) whose structure is expected to be rather
soft. Consequently, the transition form factor to the latter resonance decreases
faster than that to its doublet partner with the scalar diquark contribution.

In the strange sector\cite{Loring:2001kvC} 't Hooft's force
accounts in a similar way for the prominent doublets of the
$\Lambda$-spectrum. At the same time instanton-induced effects are
found to be significantly weaker in the $\Sigma$-spectrum, thus
explaining the fact that no clear experimental indications of parity
doublets are observed in this sector.

The contributions of K.~Kretzschmar and H.~R.~Petry are gratefully
acknowledged.



\begin{thebibliography}{99}
\bibitem{PDG00} 
Particle Data Group, \Journal{\EPJC}{15}{1}{2000}.


\bibitem{Glozman:2000tk}
L.~Y.~Glozman,
\Journal{\PLB}{475}{329}{2000}

\bibitem{Cohen:2001gb}
T.~D.~Cohen and L.~Y.~Glozman,
hep-ph/0102206.

\bibitem{Kirchbach:2001hj}
M.~Kirchbach,
\Journal{\NPA}{689}{157}{2001}


\bibitem{Loring:2001kvA}
U.~L\"oring, K.~Kretzschmar, B.~Ch.~Metsch and H.~R.~Petry,
Eur. Phys. J. A \textbf{10}, 309 (2001).


\bibitem{Loring:2001kvB}
U.~ L\"oring, B.~Ch.~Metsch,  H.~R.~Petry,
Eur. Phys. J. A \textbf{10}, 395 (2001).

\bibitem{Loring:2001kvC}
U.~ L\"oring, B.~Ch.~Metsch,  H.~R.~Petry,
Eur. Phys. J. A \textbf{10}, 447 (2001).







\end{thebibliography}
\end{document}